\documentclass[aps,prl,reprint,twocolumn,superscriptaddress,nofootinbib]{revtex4-2}
\usepackage{amsmath,amssymb,bm}
\usepackage{graphicx}
\usepackage{microtype}
\usepackage{siunitx}
\usepackage{hyperref}
\hypersetup{colorlinks=false,pdfborder={0 0 0}}
\newcommand{\Mseg}{M_{\rm seg}}
\newcommand{\Ksig}{K_{\sigma}}
\begin{document}

\title{Microscopic origin of the Baumg\"artel--Schausberger--Winter Relaxation Spectrum in Polymer Melts and Particle Rafts}

\author{Dario Nichetti}
\email{dario.nichetti@rheonicsrl.com}
\affiliation{Rheonic Lab, Via Quadelle 2C, Castelleone (CR)
26012, Italy}

\author{H. Henning Winter}
\email{winter@umass.edu}
\affiliation{Department of Chemical \& Biomolecular Engineering and Department of Polymer Science \& Engineering, University of Massachusetts Amherst, Amherst, Massachusetts 01003, USA}

\author{Alessio Zaccone}
\email{alessio.zaccone@unimi.it}
\affiliation{Department of Physics ``A. Pontremoli'', University of Milan, Via Celoria 16, 20133 Milan, Italy}
% Institutional affiliations to be inserted before submission.

\date{\today}

\begin{abstract}
Entangled polymer melts exhibit the robust two-branch Baumg\"artel--Schausberger--Winter (BSW) relaxation spectrum, while related spectra occur in nonpolymeric monodisperse disordered systems. In spite of the successful application of BSW to many different materials, a molecular derivation of these spectra is lacking. We construct a molecular theory in which a chain segment moves relative to a screened, dynamically responding multichain environment. Gaussian-chain preaveraging gives $M_{\rm seg}(\Delta m)\sim(\Delta m)^{-1/2}$, hence $\lambda_p\sim p^{3/2}$ and, after stress projection, $H(\tau)\sim\tau^{-2/3}$. Independently, longitudinal primitive-path diffusion gives contour-length fluctuations with $H(\tau)\sim\tau^{1/4}$. A molecular-weight-constrained implementation is tested simultaneously against experimental $G'(\omega)$ and $G''(\omega)$ data for four monodisperse polybutadiene (PBD) melts, without fitting spectral exponents or individual modal weights. The resulting BSW spectrum exhibits a continuous transfer from the fast cooperative to the slow constraint-renewal cascade before a finite-chain terminal edge. A common two-sector caged dynamics then connects polymers to particle rafts without assuming identical microscopic mechanisms.
\end{abstract}

\maketitle

Entangled-polymer viscoelasticity is conventionally described in terms of Rouse-like internal modes, $\lambda(q)\propto q^2$ \cite{Rouse1953}, together with long-time relaxation by escape from the confining tube \cite{deGennes1971,DoiEdwards}. Modern tube theories incorporate contour-length fluctuations (CLF), constraint release, and longitudinal relaxation and can reproduce the measured dynamic moduli with considerable accuracy \cite{Milner1998,Likhtman2002,Graham2003}. A different question, however, remains open: why does the underlying relaxation spectrum itself display such a simple and reproducible form?

Nearly monodisperse melts exhibit two algebraic relaxation cascades joined by a smooth minimum, the Baumg\"artel--Schausberger--Winter (BSW) spectrum \cite{BSW1990,Baumgaertel1992}. Typical polymer exponents are $n_E\simeq0.23$ and $n_R\simeq0.67$ \cite{Friedrich2008}. The significance of this form is not only the existence of two limiting power laws, but their continuous transfer of dominance through one spectrum. Moreover, closely related spectra have recently been reported in particle-laden interfaces \cite{Winter2026,Oliveira2026}, where the microscopic constituents and constraints are entirely different. This raises a broader fundamental question: namely, whether the BSW exponents and their smooth crossover are consequences of a common class of stress-bearing dynamics, rather than being material-specific fitting parameters.

We address these questions by constructing the relaxation operator before introducing any target spectrum. For a dense, flexible, monodisperse melt at $T>T_g$ and $M\gg M_e$, the basic coordinate is the motion of a chain segment relative to its dynamically constraining multichain environment. Coarse graining connectivity, incompressibility, and topological constraints \cite{DoiEdwards} produces a screened nonlocal cooperative mobility and a $q^{3/2}$ relaxation law. Independently, longitudinal motion along the primitive path generates the CLF $t^{1/4}$ dynamics. Stress projection then converts these two molecular sectors into the BSW exponents, while their coexistence produces the smooth crossover between the fast and slow cascades. We finally test the resulting operator against a four-molecular-weight PBD series and ask which part of the construction can survive, at coarse-grained level, in particle rafts.

\emph{Relative chain--environment dynamics.---}
The relevant short-scale coordinate is a displacement relative to that environment,
\begin{equation}
\bm u_{\rm rel}(m,t)=\bm u_{\rm ch}(m,t)-\bm u_{\rm env}(m,t).
\label{eq:urel}
\end{equation}
A uniform translation shifts both terms equally and costs no energy; a relative chain--environment displacement is not a symmetry and can acquire a finite restoring coefficient after coarse graining the multichain constraints.

The transverse kernel can be defined unambiguously from the quadratic free energy of the relative molecular displacement,
\begin{equation}
F_{\rm rel}=\frac12\int_{\bm k}u_T(-\bm k)K_T(k)u_T(\bm k).
\label{eq:KTdef}
\end{equation}
Here $K_T=\chi_T^{-1}$ is the transverse stiffness of chain motion relative to its environment. Isotropy, inversion symmetry, and finite-range noncritical coarse graining give the standard analytic expansion \cite{ChaikinLubensky1995,BaggioliZaccone2022,Zaccone2023}
\begin{equation}
K_T(k)=K_0+K_2k^2+O(k^4),\qquad \xi_d^2=K_2/K_0.
\label{eq:KT}
\end{equation}
For an absolute displacement translational invariance gives $K_0=0$; for the relative coordinate $K_0>0$ is allowed. The inverse kernel is therefore screened,
\begin{equation}
G_d(r)\propto \frac{e^{-r/\xi_d}}{r},
\label{eq:yukawa}
\end{equation}
so a local perturbation is communicated through surrounding chains over the finite dynamic length $\xi_d$.

This defines \emph{screened nonlocal cooperative segmental mobility} (SNLCSM): a force at $m'$ perturbs the shared environment and contributes to motion at $m$:
\begin{equation}
\partial_t\bm u(m,t)=\int dm'\,\Mseg(m-m')\,\bm f(m',t)+\bm\eta(m,t).
\label{eq:snlcsm}
\end{equation}
The nonlocality is therefore mediated by the responding environment, not inserted as a phenomenological memory kernel.

For two Kuhn segments separated by $\Delta m$ contour steps, Gaussian-chain statistics gives
\begin{equation}
P(\bm r|\Delta m)=\left(\frac{3}{2\pi b^2\Delta m}\right)^{3/2}
\exp\!\left[-\frac{3r^2}{2b^2\Delta m}\right].
\label{eq:gauss}
\end{equation}
Preaveraging Eq.~\eqref{eq:yukawa} with Eq.~\eqref{eq:gauss} yields
\begin{align}
\Mseg(\Delta m)&=\frac{M_0\sqrt3}{(2\pi)^{3/2}b\sqrt{\Delta m}}
\left[1-\sqrt\pi\,x e^{x^2}\operatorname{erfc}(x)\right],
\label{eq:Mseg}\\[-2pt]
x&=\frac{b\sqrt{\Delta m}}{\sqrt6\,\xi_d}.\nonumber
\end{align}
For $b\sqrt{\Delta m}\ll\xi_d$, $\Mseg\sim(\Delta m)^{-1/2}$ and $\widetilde M_{\rm seg}(q)\sim q^{-1/2}$, fixed by screened three-dimensional propagation and Gaussian-chain geometry.

The entropic chain free energy independently provides a $q^2$ elastic force, $f_q=-k_cq^2u_q$. Equation~\eqref{eq:snlcsm} then gives
\begin{equation}
\lambda_{\rm coop}(q)=k_cq^2\widetilde M_{\rm seg}(q)\sim q^{3/2}.
\label{eq:lambda32}
\end{equation}
At scales beyond the screening crossover the mobility becomes effectively local and the ordinary Rouse law $\lambda\sim q^2$ is recovered. This screened cooperative sector therefore controls the fast stress-bearing modes; the long-time branch must arise from a distinct molecular coordinate.

\emph{Independent longitudinal and terminal dynamics.---}
Stored contour length equilibrates along the primitive path through a distinct longitudinal diffusive sector,
\begin{equation}
\lambda_{\parallel}(q)=D_{\parallel}q^2,
\qquad
\langle\Delta u^2(t)\rangle\propto t^{1/2}.
\label{eq:long}
\end{equation}
The end-exploration distance therefore grows as $t^{1/4}$, giving CLF \cite{Milner1998,Likhtman2002}; reptation and thermal constraint release complete terminal tube renewal. The $3/2$ cooperative law and $1/4$ end-fluctuation law are thus independent dynamical results.

\emph{Stress projection and observable response.---}
For eigenmodes $\bm v_\alpha$ with rates $\lambda_\alpha$, the shear-stress observable fixes the modal weights
\begin{equation}
g_\alpha=|\langle\sigma_{xy}|\bm v_\alpha\rangle|^2,
\qquad
G(t)=\sum_\alpha g_\alpha e^{-\lambda_\alpha t}.
\label{eq:stressproj}
\end{equation}
For cooperative Gaussian-chain modes, $\sigma_{xy}\propto\sum_p q_p^2X_{p,x}X_{p,y}$ and equipartition gives $g_p\sim p^0$; modal weights are therefore fixed by stress projection.

For comparison with experiment we use a reduced operator whose internal rate crosses from Rouse to SNLCSM scaling,
\begin{equation}
\lambda_{\parallel}(p)=A_{\parallel}p^2,\qquad
\lambda_u(p)=A_u\frac{p^2}{\sqrt{1+p/p_x}},
\label{eq:reducedrates}
\end{equation}
while $\rho(p)=\rho_0p/(p+p_c)$ couples the sectors. Diagonalization generates the hybrid rates and stress overlaps, and one complex response gives both moduli,
\begin{equation}
G^*(\omega)=G'(\omega)+iG''(\omega)
=\sum_\alpha g_\alpha\frac{i\omega}{\lambda_\alpha+i\omega}.
\label{eq:Gstar}
\end{equation}
Individual modal weights, rates, and asymptotic exponents are not fit variables. Thus the reduced implementation tests how the independently derived sectors combine in the observable response rather than reconstructing a free relaxation spectrum. Because the same operator must generate both moduli, residual mismatches constrain the reduced closure rather than being absorbed into independent spectral shapes.

\begin{figure*}[t]
\centering
\includegraphics[width=0.82\textwidth]{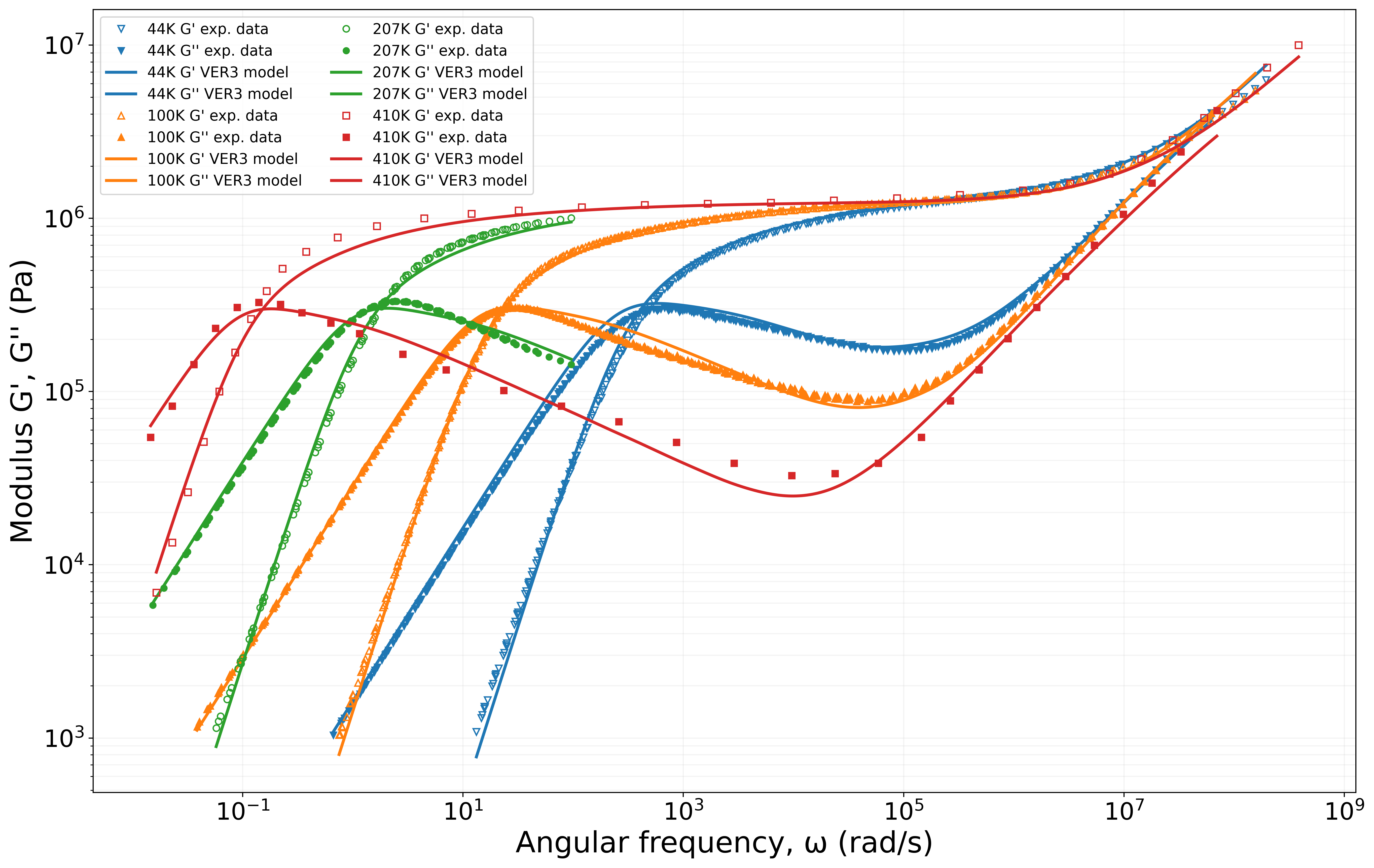}
\caption{Molecular-weight-constrained fit of $G'$ and $G''$ for four monodisperse PBD melts, $M=44$, 100, 207, and 410 kg mol$^{-1}$ \cite{Liu2006}. One reduced operator architecture with common molecular-weight laws generates both moduli; spectral exponents and individual modal weights are not fitted.}
\label{fig:pbd}
\end{figure*}

\begin{figure*}[t]
\begin{minipage}[t]{0.48\textwidth}
\centering
\includegraphics[width=0.92\linewidth]{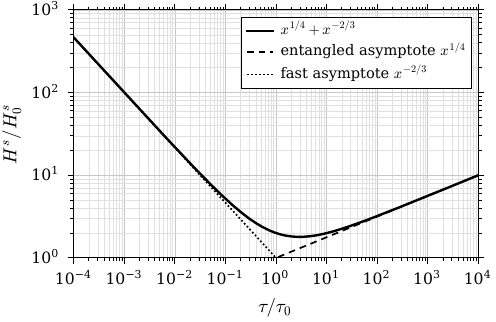}
\caption{Reduced spectrum $H^s/H_0^s=x^{1/4}+x^{-2/3}$. The minimum marks the continuous transfer of dominance between the two cascades.}
\label{fig:spectrum}
\end{minipage}\hfill
\begin{minipage}[t]{0.48\textwidth}
\centering
\includegraphics[width=0.84\linewidth]{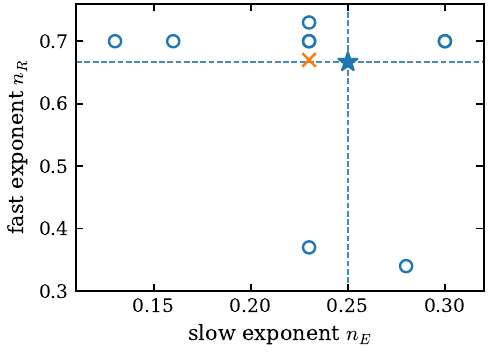}
\caption{Exponent map. Open circles: nine 2dBSW sets \cite{Oliveira2026}; star: $(1/4,2/3)$; cross: polymer BSW $\approx(0.23,0.67)$ \cite{Friedrich2008}.}
\label{fig:expmap}
\end{minipage}
\end{figure*}

Figure~\ref{fig:pbd} tests transferability across four monodisperse PBD melts \cite{Liu2006}. One operator architecture generates both moduli under common molecular-weight laws. The global description captures the terminal-to-high-frequency evolution; residual deviations expose limitations of the simplest reduced fast-sector closure (Supplemental Material).

\emph{Emergent relaxation spectrum.---}
Only after the molecular dynamics and stress projections are fixed do we introduce the continuous spectrum,
\begin{align}
G(t)&=\int_0^\infty H(\tau)e^{-t/\tau}\,d\ln\tau,\nonumber\\
H(\tau)&=\sum_\alpha g_\alpha\delta(\ln\tau-\ln\tau_\alpha).
\label{eq:Hdef}
\end{align}
If $\tau_p=A p^{-\alpha}$ and $g_p=Bp^{-\beta}$ in a scaling window, then
\begin{equation}
H(\tau)\propto \tau^{(\beta-1)/\alpha}.
\label{eq:modalmap}
\end{equation}
Thus the cooperative result $\alpha=3/2$, together with $\beta=0$ from the stress operator, gives
\begin{equation}
H_R(\tau)\sim\tau^{-2/3}.
\label{eq:fastH}
\end{equation}
Independently, the CLF end-retraction law leads to the rate density $K_{\rm CLF}(\epsilon)\propto\epsilon^{-5/4}$ and hence
\begin{equation}
H_E(\tau)\sim\tau^{1/4}.
\label{eq:slowH}
\end{equation}
Reptation and Rouse-like sectors add subleading terminal and crossover contributions. When cooperative and CLF terms dominate on opposite sides of the minimum,
\begin{equation}
H^s(\tau)=H_E^s\!\left[\left(\frac{\tau}{\tau_E}\right)^{1/4}
+\left(\frac{\tau}{\tau_R}\right)^{-2/3}\right],
\label{eq:winter}
\end{equation}
The two sectors coexist through the crossover: cooperative modes dominate at short $\tau$, while CLF modes progressively take over at long $\tau$. Their sum therefore yields a smooth transfer of stress relaxation rather than a matching of separately prescribed spectra. Equation~\eqref{eq:winter} is the BSW form obtained at the end of the construction, predicting $n_E=1/4$ and $n_R=2/3$ without fitting either exponent.

Self-similarity cannot extend to arbitrarily long times: the spectrum must ultimately cross over to terminal relaxation on the scale of the tube-renewal or disengagement time $\tau_d$. The detailed shape of this terminal edge is not fixed by the present derivation and may be statistically broadened, consistent with earlier rheological evidence that an abrupt BSW termination can be overly restrictive~\cite{Winter2013}. This terminal cutoff is distinct from the fast-to-slow crossover: the latter results from the continuous coexistence and transfer of dominance between the two relaxation cascades. Whether the statistical terminal width represents information beyond that already contained in the ensemble-averaged stress-weighted eigenvalue spectrum remains an open question of the present molecular theory.

The prediction is close to polymer BSW values $n_E\simeq0.23$, $n_R\simeq0.67$ \cite{BSW1990,Friedrich2008}. Nine reported 2dBSW sets also lie nearby overall \cite{Winter2026,Oliveira2026}, with two clear low-$n_R$ exceptions (Fig.~\ref{fig:expmap}). The sequence of fast relaxation, smooth takeover by slow modes, and terminal loss of stress memory therefore suggests a broader comparison beyond polymers.

\emph{Common caged-medium description.---}
The interfacial data suggest a common coarse-grained structure: fast displacement $u_T$ relative to a transient cage, coupled to a slower cage/constraint field $c$ \cite{Winter2026,Oliveira2026},
\begin{equation}
\partial_t\bm\Psi=-\hat L\bm\Psi+\bm\eta,\quad
\bm\Psi=(u_T,c),\quad
\hat L=\begin{pmatrix}\hat M_T\hat K_T&\hat C\\\hat C^\dagger&\hat R_c\end{pmatrix}.
\label{eq:cage}
\end{equation}
In polymers the fast block is SNLCSM and $c$ is tube memory renewed by CLF, reptation, and constraint release. In particle rafts the two sectors are localized cage rearrangements and collective contact-network restructuring. At this level the analogy is dynamical rather than structural: fast motion occurs relative to transient constraints, while slower evolution renews those constraints and eventually erases stress memory. Their terminal scale is the decorrelation lifetime of the stress-bearing cage/network (or connected cluster), not a tube-disengagement time. The microscopic operators need not be identical.

The common statement is instead spectral. For any linear dynamics define the stress-weighted rate density
\begin{equation}
\Ksig(\lambda)=\sum_\alpha g_\alpha\delta(\lambda-\lambda_\alpha),\qquad
H(\tau)=\frac{1}{\tau}\Ksig\!\left(\frac{1}{\tau}\right).
\label{eq:weighted}
\end{equation}
The same two-branch spectrum follows when the stress-projected fast and slow blocks fall into the classes
\begin{equation}
K_\sigma^{(f)}(\lambda)\sim\lambda^{-1/3},\qquad
K_\sigma^{(s)}(\lambda)\sim\lambda^{-5/4},
\label{eq:classes}
\end{equation}
mapping to $H_f\sim\tau^{-2/3}$ and $H_s\sim\tau^{1/4}$. The polymer calculation derives both classes; for rafts this remains a testable hypothesis. A cage/network Hessian and mobility-weighted operator can be constructed from configurations and projected onto interfacial shear stress \cite{BaggioliZaccone2022,Zaccone2023}. Agreement would identify a common \emph{caged dynamical class}, not common microscopic mechanisms.

\emph{Conclusion.---}
The Baumg\"artel--Schausberger--Winter (BSW) spectrum emerges here from two physically distinct but continuously overlapping relaxation cascades. Screened cooperative segmental motion controls the fast side of the spectrum, while longitudinal motion along the primitive path and the associated contour-length fluctuations progressively dominate the slow side. Their coexistence produces a smooth transfer of stress relaxation through the spectral minimum, rather than a matching of separately prescribed fast and slow branches. Finite molecular architecture establishes the characteristic terminal scale and the eventual crossover to Maxwell relaxation, while the statistical structure of this terminal crossover is not determined by the present derivation and may reflect the stochastic character of terminal molecular relaxation.

The same physical organization suggests a broader interpretation of related spectra observed in particle rafts. In both systems, stress relaxation can be viewed as the combination of fast motion relative to a transient cage, slower restructuring or renewal of that cage, and a finite terminal lifetime of the stress-bearing constraint network. The microscopic mechanisms need not be identical; what can remain common is the stress-weighted spectral structure generated by these two dynamical sectors. This provides a possible physical basis for the recurrence of Baumg\"artel--Schausberger--Winter-like spectra across otherwise very different disordered materials.

\bibliographystyle{apsrev4-2}
\bibliography{references_Winter}

\end{document}